\documentclass[a4paper,twoside,12pt]{article}
\usepackage{epsfig,rotating,physics}
\usepackage{cite}
\usepackage{a4p}
\usepackage{amssymb}
\usepackage{pennames}
\usepackage{l3_title}
\usepackage{ifthen,Lep}
\newcommand {\AcPP}   {Acta Phys. Polonica\ }
\newcommand {\EPJ}    {Eur. Phys. J.\ }
\newcommand {\SovJNP} {Sov. J. Nucl. Phys.\ }
\journalname{Phys. Lett. B}
\preprint{99-050}
\newlength{\capwidth}
\newcommand{\PZ}{\ensuremath{\mathrm{Z}}}

\newcommand{\LUBOEI}{{\scshape luboei}}
\newcommand{\PARJ}[1]{{\scshape parj(#1)}}
\newcommand{\JETSET}{{\scshape jetset}}

\newcommand{\HERWIG}{{\scshape herwig}}
\newcommand{\VNI}{{\scshape vni}}
\newcommand{\LEP}{{\scshape lep}}

\newcommand{\Lthree}{{\scshape l}{\small 3}}

\newcommand{\GEANT}{{\scshape geant}{\small 3}}
\newcommand{\GHEISHA}{{\scshape gheisha}}
\newcommand{\TEC}{{\scshape tec}}
\newcommand{\SMD}{{\scshape smd}}

\newcommand{\Qlong}{\ensuremath{Q_\mathrm{L}}}
\newcommand{\Qside}{\ensuremath{Q_\mathrm{side}}}
\newcommand{\Qout}{\ensuremath{Q_\mathrm{out}}}

\newcommand{\Qslong}{\ensuremath{Q^2_\mathrm{L}}}
\newcommand{\Qsside}{\ensuremath{Q^2_\mathrm{side}}}
\newcommand{\Qsout}{\ensuremath{Q^2_\mathrm{out}}}

\newcommand{\Rtrans}{\ensuremath{R_\mathrm{T}}}
\newcommand{\Rlong}{\ensuremath{R_\mathrm{L}}}
\newcommand{\Rside}{\ensuremath{R_\mathrm{side}}}
\newcommand{\Rout}{\ensuremath{R_\mathrm{out}}}

\newcommand{\Rslong}{\ensuremath{R^2_\mathrm{L}}}
\newcommand{\Rsside}{\ensuremath{R^2_\mathrm{side}}}
\newcommand{\Rsout}{\ensuremath{R^2_\mathrm{out}}}
\def\th{\ifmath{{\mathrm{th}}}}
\def\thr{\ifmath{{\mathrm{thr}}}}

\def\mix{\ifmath{{\mathrm{mix}}}}
\def\noBE{\ifmath{{\mathrm{noBE}}}}
\def\MC{\ifmath{{\mathrm{MC}}}}
\def\gen{\ifmath{{\mathrm{gen}}}}
\def\det{\ifmath{{\mathrm{det}}}}

\def\NDF{\ifmath{{\mathrm{NDF}}}}

\begin{document}
\begin{titlepage}
\title{Measurement of \\ an  Elongation of the Pion Source \\
           in Z~Decays 
                 \\[-1cm]}

\author{The L3 Collaboration}

\begin{abstract}\noindent
We measure Bose-Einstein correlations between like-sign charged 
pion 
pairs in hadronic \PZ~decays with the \Lthree\ detector at \LEP.
The analysis is performed in three dimensions 
in the longitudinal center-of-mass system.
The pion source is found to be elongated along the thrust axis with a ratio
of transverse to longitudinal radius of  $0.81\pm 0.02 ^{+0.03}_{-0.19}$.
\\[1mm]
\end{abstract}

%
\submitted

\end{titlepage}

\newpage
\setcounter{page}{1}

\section{Introduction}
It has long been realized that the shape and size in space-time of
a source of pions can be
determined, as a consequence of the interference of identical bosons,
from the shape and size of the correlation function of two
identical pions in energy-momentum space\cite{Gold60,Boal90,Baym}.
The space-time shape of a source during
hadronization is important experimental
information on QCD in a sector where perturbative methods are not 
applicable.

The form of the correlation function in more than one dimension
has been a major subject of theoretical study in recent 
years\cite{ref2,csor97,ref4,ref5,ref6,ref7,ref8}.
In Monte Carlo generators, 
spherical symmetry is usually assumed\cite{Lonn95,jadach,wit,kart,geiger}, while elongation
can be expected when a string-like shape is maintained\cite{sharka,ref2,csor97}.    
Experimentally,  detailed three-dimensional analyses have been done only
for heavy ion collisions~\cite{heavy1,heavy2} and for hadron-hadron
collisions\cite{na22}. 
While the volume of the pion emission region appears to be approximately
spherical for heavy ion collisions, a clear elongation is observed in
hadron-hadron collisions. 
Analyses in \Pep\Pem\ collisions have generally been limited to one dimension~\cite{refos}.
An exception is a preliminary analysis\cite{ref3},
which indicates an elongation in \Pep\Pem\ collisions.

Recently there has been a revival of interest in Bose-Einstein (BE) correlations  
of particles produced in the hadronization of energetic quarks,
mainly due to its possible impact on the measurements of
the \PW\ mass~\cite{Lonn95,jadach,wit,kart,sharka,geiger} in the four-jet channel
\Pep\Pem$\rightarrow$\PWp\PWm$\rightarrow$\,\Pq\Paq\Pq\Paq.
The experimentally observed radius of hadron emission is of order 1 fm, 
about an order of magnitude larger than 
the distance between the \PWp\PWm\  decay vertices at present energies.
The actual overlap in \PWp\PWm$\rightarrow$\,\Pq\Paq\Pq\Paq\ decays is of course determined by the
(possibly non-spherical) configuration of the decay partons of the W's in all space-time dimensions 
during hadronization. 
A better understanding of BE correlations in $\Pep\Pem\rightarrow\PZ\rightarrow\Pq\Paq$,
using high statistics \LEP\ data,
will contribute to understanding the BE effect in \PWp\PWm.

In this letter, two-particle correlations of like-charged pions are  
studied in hadronic \PZ-decay
as a function of three components of the four-momentum difference $Q$.
Effects due to hard gluon radiation, \PZ-decay into heavy quarks or the
dependence on the transverse mass of the particles 
are not considered.

\section{Data}
\subsection{Charged-Particle and Event Selection}
The data used in the analysis were collected by the \Lthree\ detector~\cite{l3}
in 1994 at a   center-of-mass energy of $\sqrt{s}\simeq 91.2$ \GeV. The data
selection uses information on charged particles
from the Time Expansion Chamber (\TEC) and the Silicon Microvertex Detector
(\SMD)~\cite{l3}.

To obtain a data sample of hadronic \PZ\ decays, we  perform an event selection 
using charged tracks.
The charged tracks are
required to have at least 40 (of 62 possible) hits in \TEC, and
the number of wires between the first and last hit is required to be at least 50.
The distance of closest approach (projected onto the transverse plane) 
of a track to the nominal interaction vertex is required to be less than 5 mm. 
The transverse momentum of a track must be greater than 100 \MeV.

In order to reduce background arising from beam-gas and beam-wall interactions
as well as from leptonic events and from two-photon interactions, we use the following criteria
\begin{equation}
  \frac{\sum_{i}\left| \vec{p}_{i}\right|}{\sqrt{s}}> 0.15, \quad
  \frac{\left| \sum_{i} p_{\parallel i}\right|}{\sum_{i}\left|\vec{p}_{i}\right|}< 0.75, \quad 
  \frac{\left| \sum_{i} \vec{p}_{\perp i}\right|}{\sum_{i}\left|\vec{p}_{i}\right|}< 0.75, \quad
  N_{\mathrm{ch}}> 4,
\end{equation}
where $\vec{p}_{i}$ is the momentum of particle $i$,  with components
$p_{\parallel i}$ parallel to the beam direction and $\vec{p}_{\perp i}$ 
in the plane perpendicular to the beam direction, 
and where the sums run over all the tracks of the event. 
$N_{\mathrm{ch}}$ is the number of charged-particle tracks.
A small remaining background of
$\tau^{+}\tau^{-}$ events in which both $\tau$ particles decay into more than one
charged particle is  removed 
by requiring the second largest angle  $\varphi_{2}$ between any two
neighboring tracks in the $R-\varphi$ plane to be less than $170^{\circ}$.
After this cut the $\tau^{+}\tau^{-}$ contamination is approximately $8\cdot10^{-5}$.

To make sure that events lie within the full acceptance of the \TEC\ and \SMD, 
we require $\left| \cos \theta_{\thr}\right| < 0.7 $,
where $\theta_{\thr}$ is the polar angle of the thrust axis determined
from charged tracks only.

A total of about one million events satisfy the selection criteria.

For the computation of the four-momentum difference $Q$, the resolution of
the angle between pairs of tracks is of
crucial importance, especially for small $Q$-values. 
For this reason we impose an additional 
cut, requiring an unambiguous polar angle measurement.
This ensures 
good resolution of variables, such as $Q$, which depend on two tracks.
Together with the previous cuts about 40\% of the tracks are rejected. 
The final data sample consists of approximately 36 million like-sign track pairs.

\begin{figure}
\begin{center}
\epsfig{figure=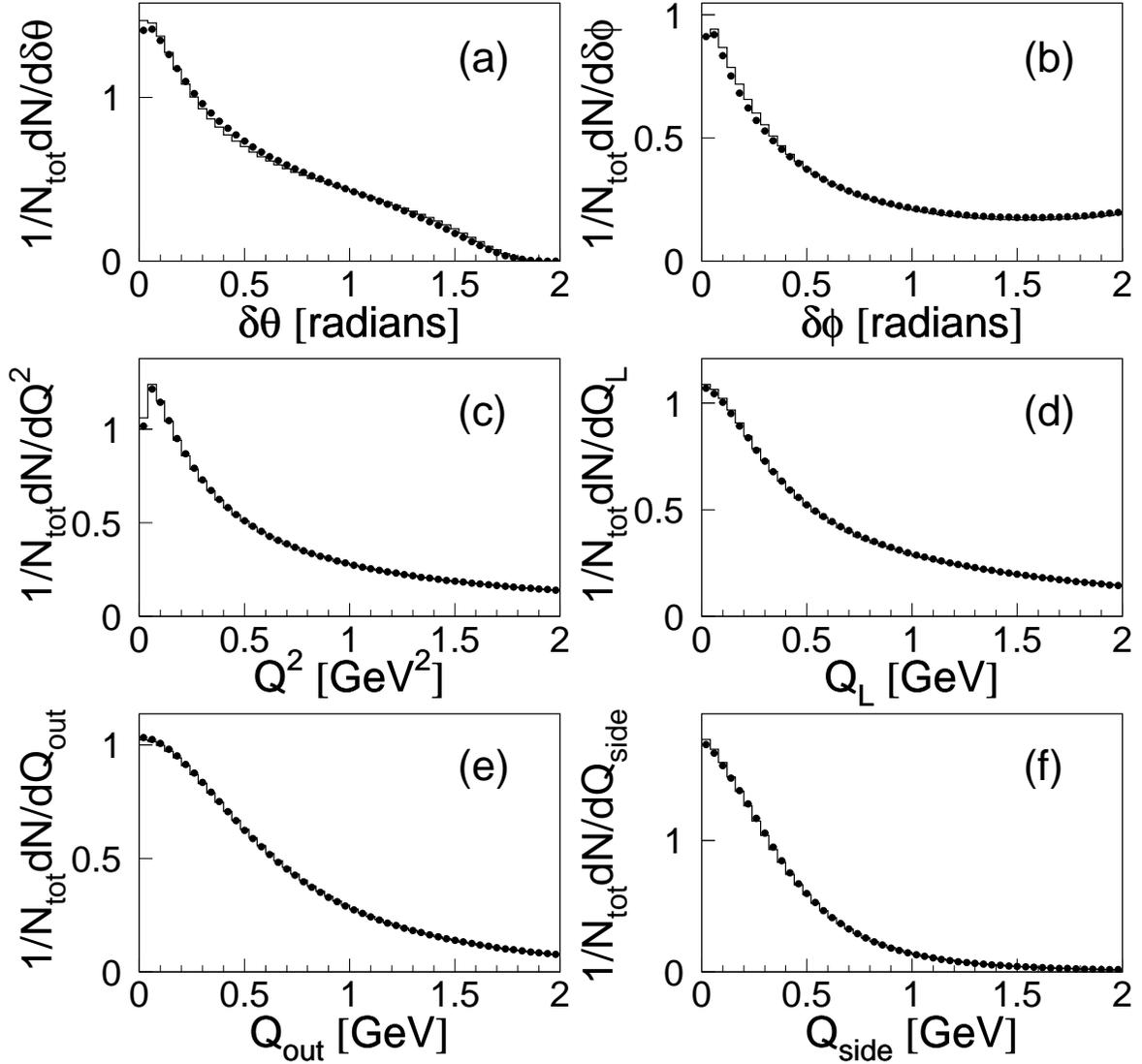, width=1.0\linewidth}
\caption{Distributions of
  (a) the difference in polar angle of pairs of tracks, $\delta\theta$,
  (b) the difference in azimuthal angle of pairs of tracks, $\delta\phi$,
  (c) the four-momentum difference squared, $Q^2$,
 and
the (d) longitudinal, (e) out, and (f) side
components of the four-momentum difference $Q$, for data (points)
compared to the predictions of \JETSET\ with BE after detector simulation (histogram).}
\label{datsimfig}
\end{center}
\end{figure}

With this selection, good agreement is obtained between data and simulation 
for the distributions of 
the differences  between pairs of tracks of the azimuthal and polar angles
with respect to the beam 
and for the distributions of $Q^{2}$ and the components of $Q$ used in this analysis.
This is shown in Fig.~\ref{datsimfig},
where the data are compared to events generated with  \JETSET~\cite{jet} (including BE effects) 
which have been passed through the \Lthree\ detector simulation program~\cite{detsim}, 
reconstructed, and subjected to track and event selection in the same way as the data.
Similar comparisons using events generated by \JETSET\ without BE effects or by \HERWIG~\cite{herw}
also show good agreement except at small values of the variables,
where the Bose-Einstein effect is important.

\section{Analysis}
\subsection{Bose-Einstein Correlation Function}
The two-particle correlation function of two particles with
four-momenta $p_{1}$ and $p_{2}$ is given by the ratio of the two-particle number density,
$\rho_2(p_{1},p_{2})$,
to the product of the two single-particle number densities, $\rho_1 (p_{1})\rho_1 (p_{2})$.
Since we are here interested only in the correlation $R_2$ due to Bose-Einstein
interference, the product of single-particle densities is replaced by 
$\rho_0(p_1,p_2)$,
the two-particle density that would occur in the absence of Bose-Einstein correlations:
\begin{equation}
  R_2(p_1,p_2)=\frac{\rho_2(p_1,p_2)}{\rho_0(p_1,p_2)}
\ \ \ .
\label{eq1}
\end{equation}
$R_2-1$ is related to the space-time particle density through 
a Fourier transform\cite{Gold60,Boal90}.

Since the mass of the two identical particles of the pair is fixed to the pion mass,
the correlation function is defined in six-dimensional momentum space.
Since Bose-Einstein correlations can be large only at small four-momentum difference
$Q=\sqrt{-(p_1-p_2)^2}$, they are often 
parametrized in this one-dimensional distance measure. There is no reason, however,
to expect the hadron source to be spherically symmetric in jet fragmentation.
This is the reason for performing a three-dimensional analysis.

\subsection{Longitudinal Center-of-Mass System}
\begin{figure}
\begin{center}
\epsfig{figure=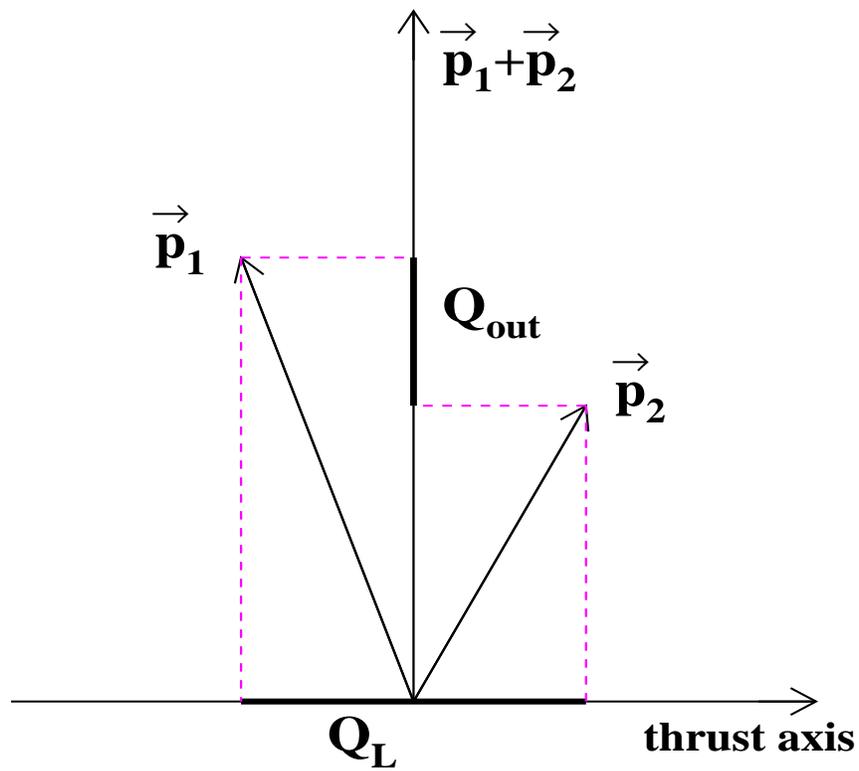, height=15cm,width=1.0\linewidth}
\caption{The longitudinal center of mass frame (LCMS) showing the projection of $Q$ on the (\Qlong-\Qout)
plane. \Qside\ is the projection of $Q$ on the axis perpendicular to this plane.}
\label{lcmsys}
\end{center}
\end{figure}

In our analysis we use the longitudinal center-of-mass system (LCMS)~\cite{ref6}. 
This is defined for each pair of particles as the system,
resulting from a boost along the thrust axis,
in which the sum of the momenta of the pair is perpendicular to the thrust axis. 
In this system, we can resolve the three-momentum difference of the pair of
particles into a longitudinal component \Qlong\ parallel to the thrust axis, 
\Qout\ along the sum of the particles' momenta (see Fig.~\ref{lcmsys})
and \Qside\ perpendicular to both \Qlong\ and \Qout. 
Then, the invariant four-momentum difference can be written as~\cite{ref6}
\begin{equation}
  Q^{2} = \Qlong^2 + \Qside^2 + \Qout^2 -(\Delta E)^2
        = \Qlong^2 + \Qside^2 + \Qout^2 (1-\beta^2),
\end{equation}
where
\begin{equation}
  \beta\equiv\frac{p_{\mathrm{out}\,1}+p_{\mathrm{out}\,2}}{E_{1}+E_{2}}\label{beta}
\end{equation}
with $p_{\mathrm{out}\,i}$ and $E_{i}\,\,(i=1,2)$ the out-component of the momentum and
the energy, respectively, of particle $i$ in the LCMS. 
The energy difference $\Delta E$ and therefore the difference in emission time of the two particles
couples only to the component \Qout.
Consequently,
\Qlong\ and \Qside\ reflect only spatial dimensions of the source, 
whereas \Qout\ reflects a mixture of spatial and temporal dimensions.
The correlation function is then parametrized in terms of $\vec{Q}=(\Qlong,\Qside,\Qout)$:
\begin{equation}
  R_{2}(\vec{Q})=\frac{\rho_2(\vec{Q})}{\rho_0(\vec{Q})}
    \ \ \ .
\label{eq2}
\end{equation}

\subsection{Determination of the Correlation Function \boldmath{$R_{2}$}}
After determining $\rho_2(\vec{Q})$, there are three steps in obtaining $R_{2}$.
The first step is to create a `reference sample', which is done by event mixing,
and from it to determine $\rho_0(\vec{Q})$. Then two corrections must be determined, one for non-BE
correlations lost in the event mixing and one for detector effects.

The reference sample, from which $\rho_0$ is determined, 
is formed by mixing particles from different data events in the following way.
First, events are rotated to a system with the $z$-axis along
the thrust axis and are stored in a `pool'. 
Then, events are randomly selected from this pool and their tracks replaced by tracks of the
same charge from other events in the pool of approximately the same multiplicity 
under the condition that no track originates from the same event.
Finally, \Qlong, \Qout\ and \Qside\ are calculated for each pair.
Used events are removed from the pool, thus preventing
any regularities in the reference sample.
From this mixed sample we obtain the particle density $\rho_\mix(\vec{Q})$.

Since this mixing procedure removes correlations other than just those
of Bose-Einstein, \eg, those from energy-momentum conservation and
from resonances, $\rho_\mix$ is then corrected for this loss by a factor $C_{\mix}$, 
which is estimated by Monte Carlo (MC) using a 
generator with no Bose-Einstein (BE) effects (\JETSET\ or \HERWIG).  
Thus, in the absence of Bose-Einstein correlations,
the corrected two-particle density is given by 
\begin{equation}
  \rho_0(\vec{Q}) = \rho_{\mix}(\vec{Q})\cdot C_{\mix}(\vec{Q})
       \qquad\mathrm{,\ where}\quad C_{\mix}(\vec{Q}) =
                          \left[\frac{\rho_2(\vec{Q})}{\rho_\mix(\vec{Q})}\right]_{\MC,\,\noBE}
\label{eqrefer}\ \ . \end{equation}

The ratio ($\rho_2/\rho_{\mix}$) 
must further be corrected for detector resolution, acceptance and efficiency 
and for particle misidentification. 
For this we use a multiplicative factor derived from Monte Carlo studies.
Since the \Lthree\ detector does not identify the hadrons, 
this factor, $C_{\det}$, is given by the ratio 
of the two-pion correlation function found from MC events at generator level
to the two-particle correlation function found using all particles after full detector simulation,
reconstruction and selection:
\begin{equation}
  C_\det(\vec{Q}) =
     \left(\frac{\rho_2(\vec{Q})}{\rho_\mathrm{mix}(\vec{Q})}\right)_{\mathrm{\kern -0.2em gen,\,pions}} 
     \,\left/\,
     \left(\frac{\rho_2(\vec{Q})}{\rho_\mathrm{mix}(\vec{Q})}\right)_{\mathrm{\kern -0.2em det,\,all}}
     \right.
   \ \ .
  \label{cdeteq}
\end{equation}
Taking all charged particles, instead of only pions,
in the generator level MC, leads to consistent results.
Combining this correction factor with (\ref{eq2}) and (\ref{eqrefer}) results in
\begin{equation}
  R_{2}(\vec{Q})=\frac{\rho_2(\vec{Q})}{\rho_{\mix}(\vec{Q})}
                     \cdot \frac{1}{C_\mix(\vec{Q})}
                     \cdot C_{\det}(\vec{Q})                 \ .
  \label{r2corr}
\end{equation}
The analysis is done in three-dimensional bins of $\vec{Q}$.  
In terms of numbers $N_{klm}$ of like-sign particle pairs
in the three-dimensional bin $k,l,m$ of \Qlong, \Qout\ and \Qside, eq.~(\ref{r2corr}) becomes
\begin{equation}
  R_{2\,klm} = \left[\frac{N_{klm}}{N_{klm}^{\mix}}\right]_\mathrm{data}
               \cdot\left[\frac{N_{klm}^{\mix}}{N_{klm}}\right]_{\MC,\,\noBE}
               \cdot\left[\frac{N_{klm}^{\gen}}{N_{klm}^{\gen,\,\mix}}
                          \cdot\frac{N_{klm}^{\det,\,\mix}}{N_{klm}^{\det}}\right]_{\MC}  \ \ .
  \label{eqcorr}
\end{equation}
Each $N_{klm}$ is normalized to the total number of pairs in the corresponding sample.

The resolution in the variables $Q_i$ 
is estimated using Monte Carlo events
to be 0.02--0.05 \GeV\ for $Q_i<0.2 \GeV$.
Given
the available statistics, we choose a bin size of 0.08 \GeV.

In our analysis, we use \JETSET\ without BE and \HERWIG\ (not having  BE)
to determine the mixing correction factor $C_\mix$
and \JETSET\ (with and without BE) as well as \HERWIG\ to determine the detector
correction factor $C_\det$.
Together with a variation of the mixing technique, the selection criteria and
the fit range, these six Monte Carlo combinations will serve to estimate
systematic errors.

\subsection{Parametrization of \boldmath{$R_{2}$}}
Assuming a Gaussian (azimuthally, but not necessarily spherically, symmetric)
shape of the source,
the following three-dimensional parametrization has been 
proposed~\cite{ref4,ref5,crosst1}:
\begin{eqnarray}
  \lefteqn{R_2(\Qlong,\Qout,\Qside) =\qquad\qquad}  \nonumber \\
       & &   \gamma \left(1+\delta \Qlong+\varepsilon
          \Qout +\xi \Qside\right) \nonumber \\
       & & \cdot \left[  1+\lambda \exp \left( -\Rlong^2\Qlong^{2}
         -\Rout^2\Qout^2-\Rside^2 \Qside^2
         +2\rho_\mathrm{L,out}\Rlong\Rout\Qlong\Qout\right)
           \vphantom{\sqrt{R^2} }
         \right] ,
  \label{eqparcross3}
\end{eqnarray}
where the factor $(1+\delta \Qlong+\varepsilon \Qout+\xi\Qside)$
takes into account possible long-range momentum correlations in the form
of a slow rise, $\gamma$ is a normalization factor close to unity and
the term between square brackets is the two-particle Bose-Einstein correlation
function associated with a Gaussian shape of the source.

By fitting the correlation function with this parametrization, one can extract
the incoherence factor $\lambda$, which measures the strength of the
correlation, and the `radii' $R_i$ ($i =$ L, out and side)
defined as $1/\sqrt{2}\sigma_i$, with $\sigma^{2}_i$ the variances of
a multi-dimensional Gaussian distribution of the source in configuration space.
$\rho_\mathrm{L,out}$ is the correlation between the longitudinal and out components of this Gaussian.
In the LCMS, the duration of particle emission only couples to the 
out-direction and only enters in the parameters \Rout\ and $\rho_\mathrm{L,out}$.     
Hence, \Rside\ can be interpreted as the transverse component of the geometric radius.
The parametrization, eq.~(\ref{eqparcross3}), assumes azimuthal symmetry of the
source, which means that the two-particle Bose-Einstein correlation function
associated with the Gaussian shape of the source, is invariant under the transformation
$\Qside\rightarrow -\Qside$. Consequently, the only possible off-diagonal
term is the $\Qlong\Qout$ term.

We first checked the method using Monte Carlo events at detector level, 
which were generated by \JETSET\ without BE, instead of data.
The correction factor $C_\mix$ was determined using \JETSET\ without BE and using \HERWIG,
while $C_\det$ was determined using \JETSET\ with BE and using \HERWIG, giving four
different combinations of correction factors.
Also we used events generated by \HERWIG\ as data, with the same two MC determinations of $C_\mix$
but with \JETSET\ with BE and \JETSET\ without BE for $C_\det$.
In all eight cases fits
gave results consistent with $\lambda=0$ ($\chi^{2}/\NDF\simeq 1$ for
$\lambda=0$), as expected in the absence of Bose-Einstein correlations.

\section{Results} 

The results of a three-dimensional fit, in the range $Q_i<1.04 \GeV$,
of the Bose-Einstein correlation
function $R_{2}$ with the parametrization of eq.~(\ref{eqparcross3})
are presented in Table~\ref{tab3d1}.
The off-diagonal term turns out to be zero within errors
($\rho_\mathrm{L,out}=-0.008\pm0.057$)
and the results given in the table
correspond to a fit with this term fixed to zero.
The values are obtained using \JETSET\ without
BE for the mixing correction and \JETSET\ with BE for the detector
correction, since the latter model is found to be the most
successful in reproducing our data in the relevant variables (Fig.~\ref{datsimfig})
and since this choice gives the lowest $\chi^2$ in the fits.

To estimate the systematic errors on the fit parameters, we examined 
four different sources, repeating the full analysis in each case.
First, we looked at the fit results obtained
with the six possible combinations of mixing and detector Monte Carlo corrections.
The systematic error from this source is taken as the RMS of these six values.
Secondly, the influence of a different mixing sample was studied by removing the 
conditions that tracks are replaced by tracks with 
the same charge and coming from events with approximately the same multiplicity. 
For each of the six Monte Carlo combinations the difference in the results between the two
mixing methods was taken as an estimate of the systematic error from this source and 
the square root of the mean of the squares of these differences taken as the systematic error
from this source.
In the same way systematic errors related to track/event selection and to the choice
of fit range were evaluated.
The analysis was repeated
with stronger and weaker selection criteria,
resulting in approximately 11\% fewer/more events and 6\% fewer/more tracks. 
The upper limit of the fit range was varied by $\pm 0.16$ \GeV\ in all $Q_i$.
As a total systematic error (second error in the table) we
add the four errors from these sources in quadrature.

For \Rlong\ the  errors from all the sources are approximately equal.
However, the systematic error on $\lambda$ and the transverse radii 
is dominated by the contribution
from the six possible MC combinations to determine $C_\det\cdot C_\mix$.
Of the two, $C_\det$ shows the larger generator dependence and accounts for most
of the systematic error.  It is approximately unity for most values of $\vec{Q}$, 
but increases to about 1.1 near $Q=0$ in the case of \JETSET\ with BE and
decreases to about 0.95 for \JETSET\ without BE and to about 0.9 for \HERWIG.
We understand this difference in behavior between \JETSET\ with BE and 
the two Monte Carlos without BE as follows.  Since $Q$ depends both on the energies of the
particles as well as on the angle between them, small $Q$ can be due to either small angle
or low energies.  In a Monte Carlo with BE the fraction of pairs at small $Q$ with small
angle will be larger than in the other Monte Carlo models.  This will lead to lower
detection efficiency and hence to larger corrections.
This part of the error is also responsible for the asymmetry in the errors 
since our best choice of MC combination does not
coincide with the average. In particular, all of the other combinations result 
in smaller values of $\Rside/\Rlong$.

\begin{figure}
\begin{center}
\epsfig{figure=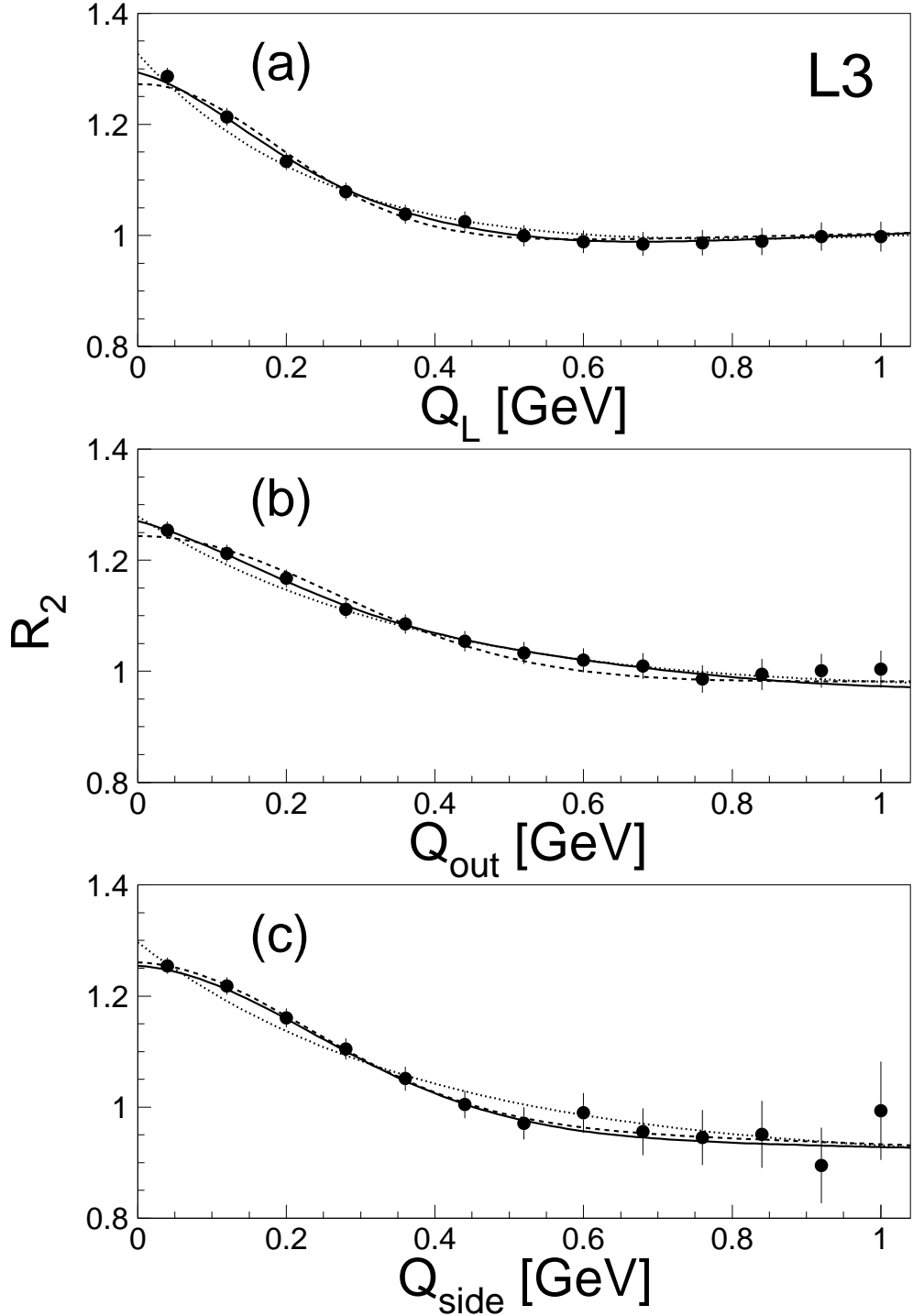, width=.8\linewidth}
\caption{Projections of $R_2$ onto the three axes \Qlong,
\Qout\ and \Qside\  using the regions up to 240 MeV
of the non-projected components.  
The mixing correction is determined using \JETSET\ without BE and the detector correction
using \JETSET\ with BE. 
The full lines correspond to projections
of the fit with the lowest-order Edgeworth expansion, the dashed lines to those of the
Gaussian fit and the dotted lines to the exponential fit.
}
\label{proj}
\end{center}
\end{figure}

\begin{figure}[t]
\begin{center}
\centering\epsfig{figure=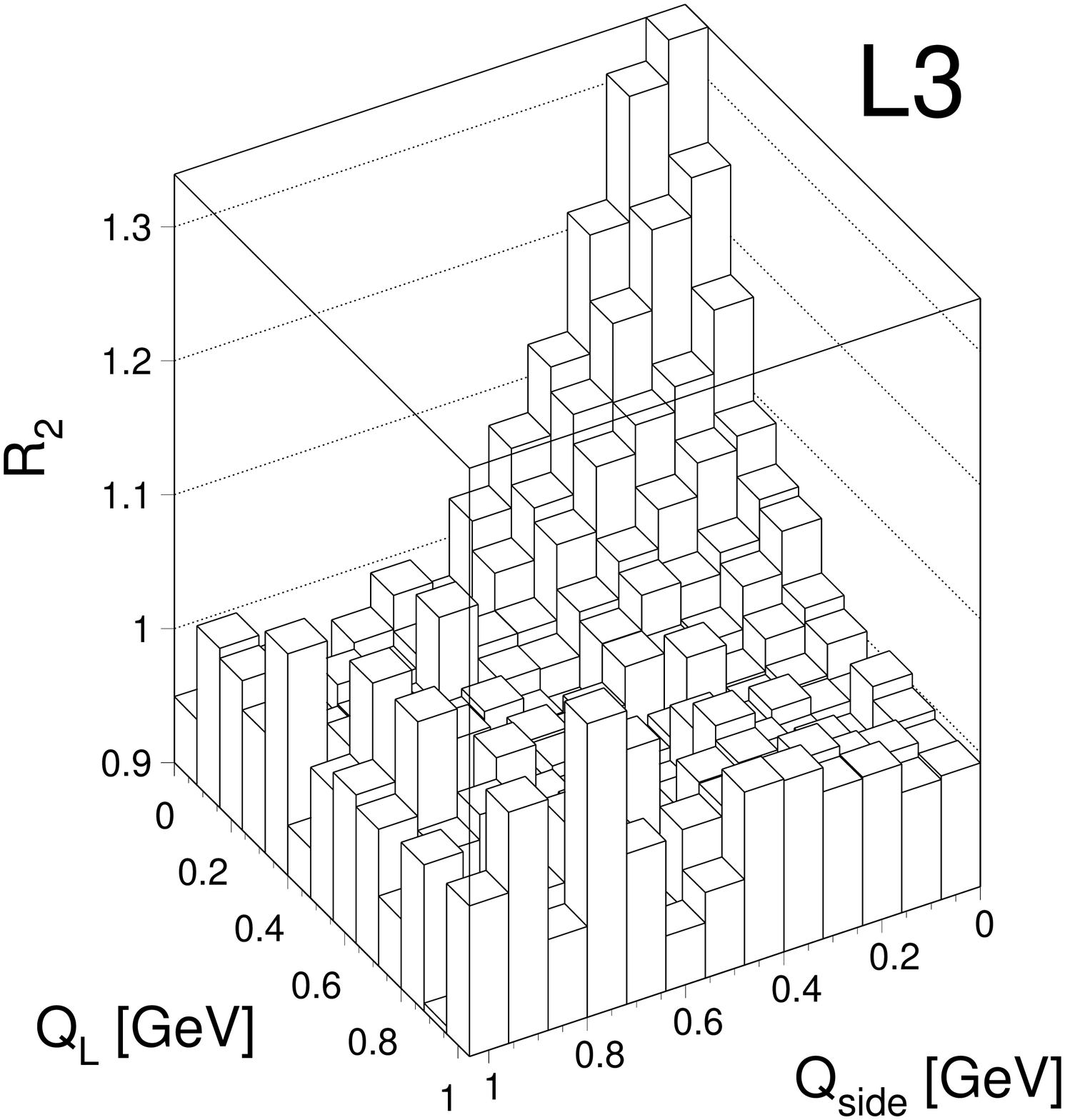, width=\linewidth}
\caption{Projection of $R_2$ onto the \Qside-\Qlong\ plane
using the regions up to 240 MeV of \Qout.
The mixing correction is determined using \JETSET\ without BE and the detector correction
using \JETSET\ with BE.}
\label{R2all}
\end{center}
\end{figure}

To study the behavior of the components of $Q$, projections 
of $R_{2}$, eq.~(\ref{eqcorr}), onto the three axes are shown in Fig.~\ref{proj},
using the regions $Q_{i}<240 \MeV$ (\ie, the first three bins)
of the non-projected components. 
The dashed curves correspond to the fit results described above.
Similarly, a projection of $R_{2}$ onto the \Qside-\Qlong\ plane is shown in Fig.~\ref{R2all}.
From Table~\ref{tab3d1} we find that \Rlong\ is larger than both \Rout\ and \Rside.
The ratios $R_\mathrm{side}/R_\mathrm{L}$ and $R_\mathrm{out}/R_\mathrm{L}$ are 5 standard
deviations less than unity.
Thus, the pion source is elongated along the thrust axis.

From the value of $\chi^2$ (see Table~\ref{tab3d1}) it appears that the
shape of the correlation function deviates from a Gaussian. 
An exponential has often been suggested as an alternative to the Gaussian.
Accordingly,
the fits were repeated with an exponential, exp$(-\sum_{i}R_{i}Q_{i})$,
replacing the Gaussian in eq.~(\ref{eqparcross3}) with $\rho_\mathrm{L,out}=0$
(dotted lines in Fig.~\ref{proj}).
Although the overall $\chi^2$ increases by typically 2-4\%, depending on the
correction combination used, the fits confirm
the elongation observed from the Gaussian fit.

A general approach to study deviations from the Gaussian\cite{edge1}, is to use
an expansion, due to Edgeworth\cite{edge2},
in terms of derivatives of the Gaussian, which are related to Hermite polynomials.
Taking only the lowest-order non-Gaussian term into account, results in
\begin{eqnarray}
  \lefteqn{R_2(\Qlong,\Qout,\Qside) =\qquad\qquad}  \nonumber \\
       & &   \gamma \left(1+\delta \Qlong+\varepsilon
          \Qout +\xi \Qside\right)        \nonumber \\
       & & \cdot\left\{ \vphantom{\left[ \frac{\kappa_\mathrm{L}}{3!}  \right] }
           1+\lambda\exp \left( -\Rslong\Qslong
               -\Rsout\Qsout-\Rsside\Qsside\right)  \right. \nonumber \\
       & & \left.
           \cdot\left[1+\frac{\kappa_\mathrm{L}}{3!}H_{3}(\Rlong\Qlong)\right]
           \cdot\left[1+\frac{\kappa_\mathrm{out}}{3!}H_{3}(\Rout\Qout)\right]
           \cdot\left[1+\frac{\kappa_\mathrm{side}}{3!}H_{3}(\Rside\Qside)\right]
           \right\} \ ,
  \label{paredge}
\end{eqnarray}
where 
$\kappa_i$ ($i=\mathrm{L,out,side}$) is the third-order cumulant moment
in the corresponding direction and 
$H_{3}(R_{i}Q_{i})\equiv (\sqrt{2}R_{i}Q_{i})^{3}-3\sqrt{2}R_{i}Q_{i}$ 
is the third-order Hermite polynomial.
Note that the second-order cumulant corresponds to the radius $R_{i}$. 
The results of the fit with
the Edgeworth expansion are shown in Table~\ref{tab3d1} and Fig.~\ref{proj} (full lines).
The value of $\chi^2/\mathrm{NDF}$ and the confidence level indicate a better fit than the Gaussian one.
The non-zero values of the $\kappa$ parameters indicate the deviation from a Gaussian. 
$\lambda$ is larger than the corresponding Gaussian $\lambda$.  
The values of the radii confirm the elongation observed from the Gaussian fit. 

For comparison we have also performed fits on a \JETSET\ generator sample
taking $R_{2}$ as
\[
   R_{2,\mathrm{JETSET}}= \frac{(N/N^\mix)_{\mathrm{JETSET\,with\,BE}}}
                               {(N/N^\mix)_{\mathrm{JETSET\,no\,BE}}}   \ .
\]
The resulting values of $\Rside/\Rlong$ and $\Rout/\Rlong$ are
 $1.08\pm 0.03$ and
 $0.79\pm 0.02$, respectively.
This value of $\Rside/\Rlong$ is larger than in the data (see Table~\ref{tab3d1}).
Thus, the standard Bose-Einstein implementation   
of \JETSET\footnote{The Bose-Einstein simulation is done by the subroutine \LUBOEI, with the 
\Lthree\ default values 
 \PARJ{92}=1.5 and \PARJ{93}=0.33 \GeV.}
fails to reproduce the experimentally observed elongation.
It is worth noting that even though there is no explicit spatial asymmetry in the \JETSET\
treatment of BE, neither \Rside\ nor \Rout\ is found to be equal to \Rlong.

Another Monte Carlo generator with Bose-Einstein simulation is \VNI\cite{vni}, 
which in its present form has been found to predict $\Rside\approx\Rlong$ \cite{geiger}, 
in contradiction to our results.

Since the difference between the two transverse components of $R$ is small
compared to that between the longitudinal component
and either of the transverse components, we check the results of the 
three-dimensional analysis by a two-dimensional analysis in \Rlong\ and
\Rtrans, where we can use intervals of 40 instead of 80 \MeV.
The out and side terms in the exponential of eq.~(\ref{eqparcross3})
are replaced by $R^2_\mathrm{T}Q^2_\mathrm{T}$, 
with $Q^2_\mathrm{T}=Q^2_\mathrm{out}+Q^2_\mathrm{side}$.
The two-dimensional fits (both Gaussian and Edgeworth) result in values of $\lambda$ consistent with
those from the three-dimensional fits and 
with values of $\Rtrans/\Rlong$ lying between the values of $\Rout/\Rlong$ and $\Rside/\Rlong$,
confirming the elongation observed in the three-dimensional fits.
As in the three-dimensional fits, the confidence level of the Gaussian fit is poor, that of the
exponential fit poorer, and that of the Edgeworth fit acceptable.
These statements are also true for analyses of \JETSET.

\section{Summary}
A sample of 1 million events of the
electron-positron annihilation process
$\Pep\Pem\rightarrow \PZ \rightarrow \mathrm{hadrons}$ at \LEP,
has been used to study
two-particle Bose-Einstein  correlations of like-charged pions.
The analysis was performed in three dimensions, as well as in two dimensions,
in the longitudinal center-of-mass system. 
Using the Gaussian parametrization of eq.~(\ref{eqparcross3}) we extracted
the incoherence factor, $\lambda$, and the radii, \Rlong, \Rout\ and \Rside.
A better fit was obtained using the Edgeworth expansion (\ref{paredge}) of the Gaussian
parametrization. The transverse radius is found to be
significantly smaller than the longitudinal radius:
\begin{center}
\renewcommand{\arraystretch}{1.32}
\begin{tabular}{c}
  $\Rside/\Rlong =  0.81\pm 0.02 ^{+0.03}_{-0.19}$ \\
\end{tabular}
\end{center}

Our measurement implies that models based on the assumption of a
spherical source are too simple.

\vskip2.0cm
\noindent
{\bf Acknowledgements}

We wish to express our gratitude to the CERN accelerator divisions for
the excellent performance of LEP. We acknowledge the effort
of the engineers and technicians who have
contributed to
the construction and maintenance of the experiment. 
Clarifying discussions with
T. Cs\"org\H{o}, O. Smirnova, \v{S}. Todorova-Nov\'a 
are gratefully acknowledged.

%
\newpage
\typeout{   }     
\typeout{Using author list for paper 174 ONLY}
\typeout{$Modified: Thu Apr  1 15:55:40 1999 by clare $}
\typeout{!!!!  This should only be used with document option a4p!!!!}
\typeout{   }
%
%
%
%
%
%

\newcount\tutecount  \tutecount=0
\def\tutenum#1{\global\advance\tutecount by 1 \xdef#1{\the\tutecount}}
\def\tute#1{$^{#1}$}
\tutenum\aachen            
\tutenum\nikhef            
\tutenum\mich              
\tutenum\lapp              
\tutenum\basel             
\tutenum\lsu               
\tutenum\beijing           
\tutenum\berlin            
\tutenum\bologna           
\tutenum\tata              
\tutenum\ne                
\tutenum\bucharest         
\tutenum\budapest          
\tutenum\mit               
\tutenum\florence          
\tutenum\cern              
\tutenum\wl                
\tutenum\geneva            
\tutenum\hefei             
\tutenum\seft              
\tutenum\lausanne          
\tutenum\lecce             
\tutenum\lyon              
\tutenum\madrid            
\tutenum\milan             
\tutenum\moscow            
\tutenum\naples            
\tutenum\cyprus            
\tutenum\nymegen           
\tutenum\caltech           
\tutenum\perugia           
\tutenum\cmu               
\tutenum\prince            
\tutenum\rome              
\tutenum\peters            
\tutenum\salerno           
\tutenum\ucsd              
\tutenum\santiago          
\tutenum\sofia             
\tutenum\korea             
\tutenum\alabama           
\tutenum\utrecht           
\tutenum\purdue            
\tutenum\psinst            
\tutenum\zeuthen           
\tutenum\eth               
\tutenum\hamburg           
\tutenum\taiwan            
\tutenum\tsinghua          
{
\parskip=0pt
\noindent
{\bf The L3 Collaboration:}
\ifx\selectfont\undefined
 \baselineskip=10.8pt
 \baselineskip\baselinestretch\baselineskip
 \normalbaselineskip\baselineskip
 \ixpt
\else
 \fontsize{9}{10.8pt}\selectfont
\fi
\medskip
\tolerance=10000
\hbadness=5000
\raggedright
\hsize=162truemm\hoffset=0mm
\def\r{\rlap,}
\noindent

M.Acciarri\r\tute\milan\
P.Achard\r\tute\geneva\ 
O.Adriani\r\tute{\florence}\ 
M.Aguilar-Benitez\r\tute\madrid\ 
J.Alcaraz\r\tute\madrid\ 
G.Alemanni\r\tute\lausanne\
J.Allaby\r\tute\cern\
A.Aloisio\r\tute\naples\ 
M.G.Alviggi\r\tute\naples\
G.Ambrosi\r\tute\geneva\
% ...
\newpage
\begin{list}{A}{\itemsep=0pt plus 0pt minus 0pt\parsep=0pt plus 0pt minus 0pt
                \topsep=0pt plus 0pt minus 0pt}
\item[\aachen]
 I. Physikalisches Institut, RWTH, D-52056 Aachen, FRG$^{\S}$\\
 III. Physikalisches Institut, RWTH, D-52056 Aachen, FRG$^{\S}$
\item[\nikhef] National Institute for High Energy Physics, NIKHEF, 
     and University of Amsterdam, NL-1009 DB Amsterdam, The Netherlands
\item[\mich] University of Michigan, Ann Arbor, MI 48109, USA
\item[\lapp] Laboratoire d'Annecy-le-Vieux de Physique des Particules, 
     LAPP,IN2P3-CNRS, BP 110, F-74941 Annecy-le-Vieux CEDEX, France
\item[\basel] Institute of Physics, University of Basel, CH-4056 Basel,
     Switzerland
\item[\lsu] Louisiana State University, Baton Rouge, LA 70803, USA
\item[\beijing] Institute of High Energy Physics, IHEP, 
  100039 Beijing, China$^{\triangle}$ 
\item[\berlin] Humboldt University, D-10099 Berlin, FRG$^{\S}$
\item[\bologna] University of Bologna and INFN-Sezione di Bologna, 
     I-40126 Bologna, Italy
\item[\tata] Tata Institute of Fundamental Research, Bombay 400 005, India
\item[\ne] Northeastern University, Boston, MA 02115, USA
\item[\bucharest] Institute of Atomic Physics and University of Bucharest,
     R-76900 Bucharest, Romania
\item[\budapest] Central Research Institute for Physics of the 
     Hungarian Academy of Sciences, H-1525 Budapest 114, Hungary$^{\ddag}$
\item[\mit] Massachusetts Institute of Technology, Cambridge, MA 02139, USA
\item[\florence] INFN Sezione di Firenze and University of Florence, 
     I-50125 Florence, Italy
\item[\cern] European Laboratory for Particle Physics, CERN, 
     CH-1211 Geneva 23, Switzerland
\item[\wl] World Laboratory, FBLJA  Project, CH-1211 Geneva 23, Switzerland
\item[\geneva] University of Geneva, CH-1211 Geneva 4, Switzerland
\item[\hefei] Chinese University of Science and Technology, USTC,
      Hefei, Anhui 230 029, China$^{\triangle}$
\item[\seft] SEFT, Research Institute for High Energy Physics, P.O. Box 9,
      SF-00014 Helsinki, Finland
\item[\lausanne] University of Lausanne, CH-1015 Lausanne, Switzerland
\item[\lecce] INFN-Sezione di Lecce and Universit\'a Degli Studi di Lecce,
     I-73100 Lecce, Italy
\item[\lyon] Institut de Physique Nucl\'eaire de Lyon, 
     IN2P3-CNRS,Universit\'e Claude Bernard, 
     F-69622 Villeurbanne, France
\item[\madrid] Centro de Investigaciones Energ{\'e}ticas, 
     Medioambientales y Tecnolog{\'\i}cas, CIEMAT, E-28040 Madrid,
     Spain${\flat}$ 
\item[\milan] INFN-Sezione di Milano, I-20133 Milan, Italy
\item[\moscow] Institute of Theoretical and Experimental Physics, ITEP, 
     Moscow, Russia
\item[\naples] INFN-Sezione di Napoli and University of Naples, 
     I-80125 Naples, Italy
\item[\cyprus] Department of Natural Sciences, University of Cyprus,
     Nicosia, Cyprus
\item[\nymegen] University of Nijmegen and NIKHEF, 
     NL-6525 ED Nijmegen, The Netherlands
\item[\caltech] California Institute of Technology, Pasadena, CA 91125, USA
\item[\perugia] INFN-Sezione di Perugia and Universit\'a Degli 
     Studi di Perugia, I-06100 Perugia, Italy   
\item[\cmu] Carnegie Mellon University, Pittsburgh, PA 15213, USA
\item[\prince] Princeton University, Princeton, NJ 08544, USA
\item[\rome] INFN-Sezione di Roma and University of Rome, ``La Sapienza",
     I-00185 Rome, Italy
\item[\peters] Nuclear Physics Institute, St. Petersburg, Russia
\item[\salerno] University and INFN, Salerno, I-84100 Salerno, Italy
\item[\ucsd] University of California, San Diego, CA 92093, USA
\item[\santiago] Dept. de Fisica de Particulas Elementales, Univ. de Santiago,
     E-15706 Santiago de Compostela, Spain
\item[\sofia] Bulgarian Academy of Sciences, Central Lab.~of 
     Mechatronics and Instrumentation, BU-1113 Sofia, Bulgaria
\item[\korea] Center for High Energy Physics, Adv.~Inst.~of Sciences
     and Technology, 305-701 Taejon,~Republic~of~{Korea}
\item[\alabama] University of Alabama, Tuscaloosa, AL 35486, USA
\item[\utrecht] Utrecht University and NIKHEF, NL-3584 CB Utrecht, 
     The Netherlands
\item[\purdue] Purdue University, West Lafayette, IN 47907, USA
\item[\psinst] Paul Scherrer Institut, PSI, CH-5232 Villigen, Switzerland
\item[\zeuthen] DESY-Institut f\"ur Hochenergiephysik, D-15738 Zeuthen, 
     FRG
\item[\eth] Eidgen\"ossische Technische Hochschule, ETH Z\"urich,
     CH-8093 Z\"urich, Switzerland
\item[\hamburg] University of Hamburg, D-22761 Hamburg, FRG
\item[\taiwan] National Central University, Chung-Li, Taiwan, China
\item[\tsinghua] Department of Physics, National Tsing Hua University,
      Taiwan, China
\item[\S]  Supported by the German Bundesministerium 
        f\"ur Bildung, Wissenschaft, Forschung und Technologie
\item[\ddag] Supported by the Hungarian OTKA fund under contract
numbers T019181, F023259 and T024011.
\item[\P] Also supported by the Hungarian OTKA fund under contract
  numbers T22238 and T026178.
\item[$\flat$] Supported also by the Comisi\'on Interministerial de Ciencia y 
        Tecnolog{\'\i}a.
\item[$\sharp$] Also supported by CONICET and Universidad Nacional de La Plata,
        CC 67, 1900 La Plata, Argentina.
\item[$\natural$] Supported by Deutscher Akademischer Austauschdienst.
\item[$\diamondsuit$] Also supported by Panjab University, Chandigarh-160014, 
        India.
\item[$\triangle$] Supported by the National Natural Science
  Foundation of China.
\end{list}
}
\vfill



%

\renewcommand{\arraystretch}{1.32}
\begin{table}[t]
\centering
\caption{Values of the fit parameters for the three-dimensional analyses of
         \Lthree\ data. The first error is statistical, the second systematic.}
\label{tab3d1} \vskip0.5cm
\small
\begin{tabular}{|c|c|c|}
\hline
parameter             & Gaussian                        & Edgeworth \\
\hline
$\lambda$             & $0.41\pm 0.01 ^{+0.02}_{-0.19}$ & $0.54\pm 0.02 ^{+0.04}_{-0.26}$ \\
\hline
\Rlong\  (fm)         & $0.74\pm 0.02 ^{+0.04}_{-0.03}$ & $0.69\pm 0.02 ^{+0.04}_{-0.03}$ \\
\Rout (fm)            & $0.53\pm 0.02 ^{+0.05}_{-0.06}$ & $0.44\pm 0.02 ^{+0.05}_{-0.06}$ \\
\Rside\ (fm)          & $0.59\pm 0.01 ^{+0.03}_{-0.13}$ & $0.56\pm 0.02 ^{+0.03}_{-0.12}$ \\
\hline
$\Rout/\Rlong$        & $0.71\pm 0.02 ^{+0.05}_{-0.08}$ & $0.65\pm 0.03^{+0.06}_{-0.09}$  \\
$\Rside/\Rlong$       & $0.80\pm 0.02 ^{+0.03}_{-0.18}$ & $0.81\pm 0.02 ^{+0.03}_{-0.19}$ \\
\hline
$\kappa_\mathrm{L}$   & --                              & $0.5\pm 0.1 ^{+0.1}_{-0.2}$ \\
$\kappa_\mathrm{out}$ & --                              & $0.8\pm 0.1\pm 0.3$ \\
$\kappa_\mathrm{side}$& --                              & $0.1\pm 0.1\pm 0.3$ \\
\hline
$\delta$              & $ 0.025\pm0.005 ^{+0.014}_{-0.015} $ & $ 0.036\pm0.007 ^{+0.012}_{-0.023} $ \\
$\epsilon$            & $ 0.005\pm0.005 ^{+0.034}_{-0.012} $ & $ 0.011\pm0.005 ^{+0.037}_{-0.012} $ \\
$\xi$                 & $-0.035\pm0.005 ^{+0.031}_{-0.024} $ & $-0.022\pm0.006 ^{+0.020}_{-0.025} $ \\
\hline
$\chi^2/\mathrm{NDF}$ & 2314/2189                       & 2220/2186 \\
C.L. (\%)             & 3.1                             & 30 \\
\hline
\end{tabular}
\end{table}
%
%
%

\end{document}